\documentclass[12pt,a4paper]{article}

  \usepackage{a4wide}
  \usepackage{latexsym}
  \usepackage{epsf}
  \usepackage{amssymb}
  \usepackage{graphicx}
  \usepackage{amsmath, cite}
  \usepackage{slashed,epsfig}
  \include{braket}
  \newcommand{\q}{\hspace{1.5mm}}
\newcommand{\sq}{\hspace{0.9mm}}

\begin{document}

\begin{flushright}
\small
UG-05-07\\
SPIN-05/30  \\
ITP-UU-05/44\\
\date \\
\normalsize
\end{flushright}

\begin{center}


\vspace{.7cm}

{\LARGE {\bf Non-Extremal D-instantons }} \\

\vspace{.7cm}

{\LARGE {\bf and}} \\

\vspace{.7cm}

{\LARGE {\bf the AdS/CFT Correspondence}} \\

\vspace{1.2cm}

\begin{center}
  Eric Bergshoeff$^1$, Andr\'es Collinucci$^1$, Andr\'e Ploegh$^1$,\\
Stefan Vandoren$^2$ and Thomas Van Riet$^1$  \\[3mm]
   {\small\slshape
  $^1$ Centre for Theoretical Physics, University of Groningen, Nijenborgh
  4,\\
9747 AG Groningen, The Netherlands \\
  {\upshape\ttfamily e.a.bergshoeff, a.g.collinucci, a.r.ploegh, t.van.riet@rug.nl}\\[3mm]
   $^2$ Institute for Theoretical Physics \emph{and} Spinoza Institute \\
  Utrecht University, 3508 TD Utrecht, The Netherlands \\
  {\upshape\ttfamily s.vandoren@phys.uu.nl}}
\end{center}
\vspace{5mm}

{\bf Abstract}

\end{center}

\begin{quotation}

\small

We investigate non-extremal D-instantons in an asymptotically
${\rm AdS_5 \times S^5}$ background and the role they play in the
${\rm AdS_5}/{\rm CFT_4}$ correspondence. We find that the
holographic dual operators of non-extremal D-instanton
configurations do not correspond to self-dual Yang-Mills
instantons, and we compute explicitly the deviation from
self-duality.

Furthermore, a class of non-extremal D-instantons yield Euclidean
axionic wormhole solutions with two asymptotic boundaries. After
Wick rotating, this provides a playground for investigating
holography in the presence of cosmological singularities in a
closed universe.

\end{quotation}

\newpage

\pagestyle{plain}

\section{Introduction}

It is by now well-established that the D-instanton of IIB string
theory \cite{Gibbons:1995vg} plays an important role in the
calculation of non-perturbative contributions to the low-energy
string effective action \cite{Green:1997tv}. Several attempts have
been made in the literature to generalize this D-instanton
solution
\cite{Kim:1996hq,Einhorn:2000ct,Einhorn:2002am,Kim:2003js,
Gutperle:2002km} to more general instanton solutions of Euclidean
IIB supergravity. In a recent work \cite{Bergshoeff:2004fq} we
showed that the extremal and non-extremal D-instantons in
(asymptotically) flat space fall under the three conjugacy classes
of $SL(2,\mathbb{R})$, the duality symmetry of IIB supergravity.
The role of the deformation parameter was played by the
determinant of the solutions' $SL(2,\mathbb{R})$ charge matrix. In
some cases the non-extremal instantons can be understood in terms
of non-extremal black holes (or $p$-branes) in one (or $p+1$)
higher dimension. In other cases, the solutions correspond to
axionic Euclidean wormhole geometries with two boundaries that
asymptote flat space at infinity
\cite{Bergshoeff:2004fq,Bergshoeff:2004pg}.

In this paper we extend our investigation of the non-extremal
D-instantons to asymptotically Anti-de Sitter spaces. While our
analysis can be done in any spacetime dimension, we will focus on
the case of ${\rm AdS_5}$. This allows us to study their dual
description, via the ${\rm AdS_5/CFT_4}$ correspondence, in the
$\mathcal{N}=4$ supersymmetric Yang-Mills theory. For extremal
D-instantons in ${\rm AdS_5 \times S^5}$, the dual description in
terms of self-dual Yang-Mills instantons has been studied in great
detail \cite{Banks:1998nr,Bianchi:1998nk,Chu:1998in,Kogan:1998re,
Dorey:1998xe,Dorey:1998qh,Dorey:1999pd,Green:2002vf}. To go beyond
extremality, this will necessitate to generalize the work of
\cite{Bergshoeff:2004fq} and construct non-extremal D-instantons
in an asymptotically ${\rm AdS_5 \times S^5}$ background. Such
solutions were also discussed in \cite{Gutperle:2002km}.  In a
different context similar solutions were discussed in
\cite{Giddings:1987cg, Myers:1988sp}.

After our construction of non-extremal D-instanton solutions in
${\rm AdS_5 \times S^5}$ we calculate the dual operators in the
${\cal N}=4$ gauge theory. Since the supergravity solution is
supported by the dilaton and axion, the dual operators are ${\rm
Tr} F^2$ and ${\rm Tr} F\tilde{F}$ respectively. We will establish
that, in contrast to the extremal case, these operators do {\it
not} satisfy the self-duality constraints for Yang-Mills
instantons. The construction of explicit non-self-dual instanton
solutions for gauge group $SU(2)$ is notoriously complicated.
However, as we will show, for gauge groups $SU(N)$ this is easier
and we will suggest that such non-self dual YM instantons are the
holographic duals of certain non-extremal D-instantons.

Part of our motivation comes from applications to cosmology. As we
will see, some of the non-extremal D-instanton solutions have
metrics that describe Euclidean wormholes in Einstein frame. These
can be Wick rotated to time-dependent backgrounds with a Big Bang
and Big Crunch singularity. A similar situation appeared in
\cite{Maldacena:2004rf}, whose authors were motivated by the
search of holographic duals of closed cosmologies. Moreover,
albeit in a slightly different context, non-extremal D-instantons
have been discussed in relation to FLRW cosmologies
\cite{Bergshoeff:2005cp}.

This paper is organized as follows. In section \ref{euclidean} we
shortly review D=10 Euclidean IIB supergravity and its relation,
via compactification  over S$^5$, to the effective D=5 field
theory. In section \ref{extremal} we construct extremal and
non-extremal instantons in an ${\rm AdS}_5$ background of this
effective D=5 field theory. By construction each solution can be
uplifted to a D=10 instanton solution in an ${\rm AdS_5 \times
S^5}$ background. In section \ref{actions} we calculate the
corresponding instanton actions. Next, in section \ref{curvature}
we calculate, via the  ${\rm AdS_5/CFT_4}$ correspondence, the
operator expressions of the corresponding Yang-Mills 1-point
functions. In section 6 we review the correspondence between
extremal D-instantons and (anti-) self-dual YM instantons. In the
same section we generalize this to the case of non-extremal
instantons and propose a possible relation between one class of
these instantons and {\it non-self-dual} YM instantons. The other
class of solutions, however, is less evident to discuss in the AdS/CFT
context, but yields interesting cosmological solution after
Wick rotation. In section \ref{conclusions} we present our
conclusions.

We have added two appendices. In appendix \ref{app_pathint}, we
present the path integral formulation of the axion-dilaton system
that leads to D-instantons in the semi-classical approximation. In
that formalism we can properly explain the `wrong' sign of the
axionic kinetic term, the necessary boundary terms in the action
and we will see how the analog of the Yang-Mills $\theta$-term
arises on the gravity side. In appendix B, we present the
D-instanton solutions in $D=3$, in which expressions can be found
in closed form.

\section{D=10 Euclidean Supergravity}\label{euclidean}

Our starting point is D=10 Euclidean IIB supergravity (see for
instance \cite{Bergshoeff:2000qu}). We only need to consider the
metric $g_{\mu\nu}$, the dilaton $\phi$, the axion $\chi$, which
is a pseudo-scalar, and the 4-form potential whose 5-form
curvature tensor $F_5$ in Euclidean space is imaginary self-dual
i.e. $\star F_5=i F_5$. Setting the fermions and the other bosonic
fields equal to zero, the equations of motion for these fields are
given by:
\begin{align}
&\partial_{\mu}(\sqrt{g}\,e^{2\phi}g^{\mu\nu}\partial_{\nu}\chi)=
0\, ,
\label{D=10_axionvergelijking}\\
&\partial_{\mu}(\sqrt{g}\,g^{\mu\nu}\partial_{\nu}\phi)+
\sqrt{g}\,e^{2\phi}g^{\mu\nu}
\partial_{\mu}\chi\partial_{\nu}\chi=0\, ,\label{D=10_dilatonvergelijking}\\
&\mathcal{R}_{\mu\nu}-\tfrac{1}{2}\partial_{\mu}\phi\partial_{\nu}\phi
+ \tfrac{1}{2}e^{2\phi}\partial_{\mu}\chi\partial_{\nu}\chi
-\tfrac{1}{6}F_{\mu}{}^{\mu_{1}\mu_{2}\mu_{3}\mu_{4}}F_{\nu\mu_{1}\mu_{2}
\mu_{3}\mu_{4}}=0\, ,\label{D=10_Einsteinvergelijking}\\
&\partial_{\mu}(\sqrt{g}\,F^{\mu\mu_{1}\mu_{2}\mu_{3}\mu_{4}})=0\,
, \label{D=10_5vormvergelijking}
\end{align}
where $g = {\rm det}\, g_{\mu\nu}$ and $\mathcal{R}_{\mu\nu}$ is
the Ricci tensor. This subsector of IIB supergravity supports
Euclidean D(-1) and D3 brane solutions. Below, we briefly summarize some
features of both solutions. Note that $\chi$ has the `wrong' sign
in front of its kinetic term. We explain this in detail in
appendix \ref{app_pathint}.

Taking the near horizon limit of the Euclidean D3-brane, the
metric becomes that of Euclidean AdS$_5\times$S$^5$ and $F_5$
takes the form of the Freund-Rubin Ansatz \cite{Freund:1980xh}. In
\emph{Poincar\'e} coordinates this is given by:
\begin{align}
&ds^2=\frac{\sqrt{Q_3}}{z^2}(dz^2+d\vec{y}_4^{\,2})+\sqrt{Q_3}d\Omega_5^2\,, \label{Poincare}\\
&F_5= -\frac{i\sq Q_3}{z^5} \sq dz \wedge dy^{0}\wedge
dy^{1}\wedge dy^{2}\wedge dy^{3}- \star\,\frac{\sq Q_3}{z^5} dz
\wedge dy^{0}\wedge dy^{1}\wedge dy^{2}\wedge dy^{3} \, ,
\end{align}
where $Q_3$ is proportional to the D3-brane charge. From this
expression we notice that the radius $l$ of the 5-sphere and of
Euclidean Anti-de Sitter space (EAdS$_5$) is related to $Q_3$ via
$l^2=\sqrt{Q_3}$.

On the other hand the solution for  the D(-1)-brane, or extremal
D-instanton, in string frame reads:
\begin{align}
& ds^2=(H_{-1})^{1/2}(d\rho^2+\rho^2d\Omega_9^2), \\
& e^{\phi}=H_{-1}\,,\\
& d\chi=\pm d(H_{-1})^{-1}\, ,
\end{align}
with the harmonic $H_{-1}$ given by:
\begin{equation}
H_{-1} = g_s + \frac{Q_{-1}}{\rho^8}\, ,
\end{equation}
where the integration constant $Q_{-1}$ is proportional to the
instanton charge and $g_s \equiv e^{\phi(\infty)}$.

Since D(-1)-branes correspond to D-instantons, and (stacked)
D3-branes are the starting point of the AdS$_5$/CFT$_4$
correspondence leading to a duality with $\mathcal{N}=4$
supersymmetric Yang-Mills theory, it is {\it a priori} natural to
consider intersections of a single D-instanton with a stack of
Euclidean D3-branes,  and, if possible, to extend this to the case
of a non-extremal D-instanton. However, we would like to consider
{\it localized} intersections as opposed to {\it delocalized}
intersections involving the extremal D-instanton, which have been
considered in \cite{Kogan:1998re,Liu:1999fc}. Due to the technical
complications with the construction of the appropriate localized
brane intersections (no explicit expressions are known, see for
instance \cite{Rajaraman:2000ws}) we will only consider instanton
solutions of the effective D=5 theory that follows after
compactification over S$^5$, in the rest of this paper. By
construction, each of these instanton solutions can be uplifted to
a D=10 instanton in an AdS$_5 \times$ S$^5$ background. What we
will not consider is the (localized) brane intersection whose
near-horizon geometry gives rise to this D=10 instanton.

The D=5 field theory is obtained as follows. We split the 10-dimensional
space into the product of two parts, one with coordinates
$x_{\mu}\, ,\ \mu=0,\ldots,4\,$, and the other part with
coordinates $y_{a}\, ,\ a =0,\ldots, 4\,$. Next, we consider the
following Ansatz:
\begin{align}
& ds^2_{10}=g_{\mu\nu}(x)dx^{\mu}dx^{\nu} + g_{ab}(y)dy^{a}dy^{b}\,,\\
& \chi=\chi(x), \quad \phi=\phi(x)\,,\\
& F_{\mu \nu \rho \sigma \delta}(x)=-\frac{i}{l}\sqrt{{\rm det}
g(x)}
\, \,{\epsilon}_{\mu \nu \rho \sigma \delta}\,,\\
&F_{abcde}(y)=-\frac{1}{l}\sqrt{{\rm det}\, g(y)}\,
{\epsilon}_{abcde}\, ,
\end{align}
where ${\epsilon}$ is the 5-dimensional Levi-Civita symbol and $l$
is a constant. With this Ansatz the self-duality and equation
(\ref{D=10_5vormvergelijking}) is satisfied. The Einstein equation
(\ref{D=10_Einsteinvergelijking}) is solved in the
$y_{a}$-directions provided $g_{ab}(y)$ is the metric of the
5-sphere S$^5$ with radius $l$. At this point we are left with the
following equations for the metric and the scalars in the
$x_{\mu}$-directions:
\begin{align}
&\partial_{\mu}(\sqrt{g}e^{2\phi}g^{\mu\nu}\partial_{\nu}\chi)=
0\,,
\label{D=5_axionvergelijking}\\
&\partial_{\mu}(\sqrt{g}g^{\mu\nu}\partial_{\nu}\phi)+
\sqrt{g}e^{2\phi}g^{\mu\nu}\partial_{\mu}\chi\partial_{\nu}
\chi=0\, ,
\label{D=5_dilatonvergelijking}\\
&\mathcal{R}_{\mu\nu}-\tfrac{1}{2}\sq\partial_{\mu}\phi\partial_{\nu}\phi
+ \tfrac{1}{2}\sq e^{2\phi}\partial_{\mu}\chi\partial_{\nu}\chi
-\tfrac{1}{3}\sq
g_{\mu\nu}\Lambda=0\,,\label{D=5_Einsteinvergelijking}
\end{align}
where $\Lambda=-12/l^2$. These equations can be derived from the
following effective action:
\begin{equation}\label{D=5_Bulkactie}
\boxed{ S=-\frac{1}{2\kappa_5^2}\int\,d^5x\,
\sqrt{g}\Bigl[\mathcal{R}-\Lambda -\tfrac{1}{2}\sq
\partial_{\mu}\phi \partial^{\mu}\phi+\tfrac{1}{2}\sq e^{2\phi}\partial_{\mu}\chi
\partial^{\mu}\chi \Bigr]\, .
\, }
\end{equation}
The 5-dimensional $\kappa_5$ is related to the 10-dimensional
$\kappa_{10}$ via: $1/\kappa_5^2=l^5\pi^3 /\kappa_{10}^2$.

In the next section we will construct instanton solutions,
corresponding to this action.

\section{Instanton Solutions}\label{extremal}

We will present the instanton solutions such that one can read of
how the $SL(2,\mathbb{R})$ symmetry acts in the space of instanton
solutions. This is done by constructing the $SL(2,\mathbb{R})$
Noether charges of the instanton, as explained in
\cite{Bergshoeff:2004fq}. The charges transform under the adjoint
of $SL(2,\mathbb{R})$ and by rewriting the integration constants
of the solution as a function of the Noether charges, the
transformation properties become clear. All the solutions will
have at least $SO(5)\times SO(6)$ symmetry, where the $SO(6)$ part
is present from a 10-dimensional point of view.  The solutions
have the property that in the limit $l \rightarrow \infty$ they
reduce to the instanton solutions in a flat background constructed
in \cite{Bergshoeff:2004fq}, where $l$ is the characteristic
radius of the asymptotically ${\rm AdS_5}$ metric.

\subsection{The Extremal Instanton}

In the extremal case we assume that the metric is that of
EAdS$_5$. This is only possible if the scalars do not contribute
to the Euclidean energy-momentum tensor. In such a case one has a
solution carrying half of the Euclidean supersymmetries
\cite{Chu:1998in}. Taking the trace of
(\ref{D=5_Einsteinvergelijking}) and using
(\ref{D=5_dilatonvergelijking}) we find:
\begin{equation}
\partial_{\mu}(\sqrt{g}g^{\mu\nu}\partial_{\nu} e^{\phi})=0.
\end{equation}
Therefore $e^{\phi}$ is a harmonic function over EAdS$_5$. For the
axion $\chi$ we then find that:
\begin{equation}
\chi=\pm\Bigl[e^{-\phi}+\text{constant}\Bigr]\, ,
\end{equation}
where the + (--) refers to an instanton (anti-instanton). In Poincar\'e coordinates, the
solution reads:
\begin{equation}\label{extremal solution}
\boxed{\begin{aligned}
ds^2&=\frac{l^2}{z^2}(dz^2+d\vec{y}_4^{\,2})\,,
  \\ e^{\phi}&=|q_{-}|H\, , \\ \chi&=
  \frac{1}{q_{-}}\Bigl(H^{-1}-q_3 \Bigr)\,.
\end{aligned}}
\end{equation}
In this notation a positive (negative) sign of $q_{-}$ means that
the solution is an (anti-) instanton. The harmonic $H$ reads
\begin{equation} \label{arbitrary-point}
  H(z,\vec{y})=\frac{g_s}{|q_{-}|}
+\frac{4\sqrt{\frac{2}{3}}}{z_{0}^3}+\frac{2\sqrt{\frac{2}{3}}(1-2\frac{f^2}{z_{0}^2})
\sqrt{1+\frac{f^2}{z_{0}^2}}}{f^3}\, ,
\end{equation}
with $f(z,\vec{y})$ defined as the following $SO(1,5)$ invariant
function:
\begin{equation}
f(z,\vec{y})=\frac{\sqrt{((z_{0}-z)^2+(\vec{y}-\vec{y_{0}})^2)\q((z_{0}+z)^2+(\vec{y}-\vec{y_{0}})^2)}}{2z}\,
.
\end{equation}
This harmonic function (\ref{arbitrary-point}) has a singularity
at ($z=z_0,\,\vec{y}=\vec{y}_0$) and this is interpreted as the
position of the D-instanton.

The $SL(2,\mathbb{R})$ charge matrix of the solution is defined by
integrating over a 4-sphere inclosing the D-instanton \footnote{We
use a slightly different normalization as compared to the
definition of the charge matrix in \cite{Bergshoeff:2004fq}. This
difference in normalization  can be traced back when one takes the
limit $l \rightarrow \infty$.}:
\begin{equation}
\mathcal{Q}=\frac{1}{2\sqrt{6}Vol(S^4)}\int_{S^4}{J_{\mu}\eta^{\mu}}\,
,
\end{equation}
where the vector $J_{\mu}$ is the $SL(2,\mathbb{R})$ Noether
current which is a $2 \times 2$ matrix, and $\eta_{\mu}$ is a unit
vector everywhere perpendicular to S$^4$. The details are
explained in \cite{Bergshoeff:2004fq}, so we immediately go to the
result, namely:
\begin{equation}\label{chargematrix1}
\mathcal{Q}=
\left(%
\begin{array}{cc}
   q_3 & i q_+ \\
   i q_- & -q_3 \\
\end{array}%
\right)\, .
\end{equation}
The number $q_{+}$ is some function of the integration constants
$q_{-}$ and $q_3$. The determinant
\begin{equation}
{\rm det}\ \mathcal{Q}=q_+q_- - q_3^2 \ ,
\end{equation}
can be positive or negative. From now on we will use the symbol
$q^2=-{\rm det} \mathcal{Q}$ in our solutions\footnote{Note that, in our notation, $q^2$ can be positive or negative.}. The determinant of this matrix
equals zero for the extremal
instanton solutions i.e. $q^2=0$. \\

Next, we consider the non-extremal instantons that have
non-vanishing $q^2$. Since the metric is $SL(2,\mathbb{R})$
invariant we expect it to get deformed with the $SL(2,\mathbb{R})$
invariant charge parameter $q^2$. The idea is to `guess' a
possible deformation for the metric and then construct the whole
solution, i.e. the dilaton and the axion. These deformations are
most easily found in \emph{radial} coordinates ($r$, $\theta_1$,
\dots, $\theta_4$):
\begin{equation}
ds^2=\frac{dr^2}{1+\frac{r^2}{l^2}}+ r^2d\Omega_{4}^2.
\end{equation}
Partial properties of these solutions were already discussed in
\cite{Gutperle:2002km, Nojiri:1999sb}. It will be necessary to
discuss the cases $q^2>0$ and $q^2<0$ separately. We refer to them
as the super- and sub-extremal instanton respectively. The
solutions will have the property that they reduce to the extremal
instanton (\ref{extremal solution}) in the limit of $q^2
\rightarrow 0$.

\subsection{The Super-Extremal Instanton: $q^2>0$}

The solution reads:
\begin{equation}\label{super_extremal solution}
\boxed{\begin{aligned}
ds^2&=\frac{dr^2}{1+\frac{r^2}{l^2}+\frac{q^2}{r^6}}+
r^2d\Omega_{4}^2, \\
e^{\phi(r)}&=\,\frac{|q_-|}{q}\sinh[qH(r)]\, ,
\\ \chi(r)&=\frac{1}{q_- }\Bigl(q \coth[qH(r)]-q_3\Bigr)\, ,
\end{aligned}}
\end{equation}
where $H(r)$ is a harmonic function on the asymptotically EAdS$_5$
space, that satisfies
\begin{equation} \label{vgl voor harmonische}
\partial_{r}H(r)=-\frac{\sqrt{24}}{g^{rr}\sqrt{g}}\,  .
\end{equation}
A compact explicit solution of this differential equation is not
known in $D=5$, whereas in $D=3$ the situation simplifies as is
shown in appendix B.  In the limit of vanishing cosmological
constant ($l \rightarrow \infty$) everything can be solved
explicitly and reduces to the results obtained in
\cite{Bergshoeff:2004fq}. The three integration constants $q,
q_-,q_3$ are related to the $SL(2,\mathbb{R})$ charge matrix as is
given in (\ref{chargematrix1}). The integration constant for $H$
gets fixed in terms of the string coupling constant $g_s$ by
defining $g_s=e^{\phi(\infty)}$. Since the derivative (\ref{vgl
voor harmonische}) is strictly negative we conclude that
$e^{\phi}>0$ for all $r$ if we choose $g_s>0$, as it should be.
The harmonic function has a singularity at $r=0$ which we identify
with the position of the instanton. This position corresponds to
$z=l$ and $\vec{y}=0$ in Poincar\'e coordinates. Since the
$SO(1,5)$ symmetry is broken by the deformation to an $SO(5)$
symmetry, there exist less bosonic collective coordinates,
resulting in the fact that the singularity (position of the
instanton) cannot be moved around in this space.

The coordinate $r$ runs from $r=0$ to $r=\infty$ and this patch
covers the whole manifold. At $r=0$ the Ricci scalar blows up
($\mathcal{R}\sim 12q^2/r^8 $) and hence there is a genuine
singularity. This is drawn suggestively in figure 1. At this
curvature singularity the harmonic has a singularity too and hence the
dilaton blows up and we cannot trust the supergravity
approximation. One might hope that string theory corrections
resolve this singularity. Nonetheless, the solution can be trusted
away from the singularity.

For this super-extremal solution there exists an interesting limit
for which one rescales the integration constants and then let
$q_{-}\rightarrow 0$. This induces a consistent truncation of the
axion \cite{Bergshoeff:2004fq}, hence this solution reduces to a
solution of a dilaton-gravity system.  One ends up with the same
metric but now the dilaton reads $\phi=\text{log}(g_s) + qH $,
which is clearly a solution of the dilaton equation $\Box\phi=0$.
This deformation of EAdS$_5$ has been studied before, for instance
in \cite{Nojiri:1999sb}.
\begin{figure}
\begin{center}
\includegraphics[scale=.6]{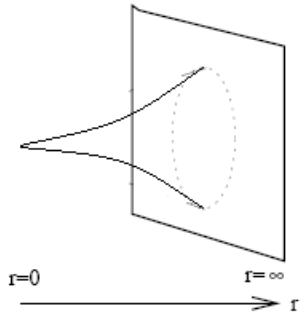}
\end{center}
\caption{\small{For the class $q^2>0$ the space looks like a
one-sided wormhole which closes at $r=0$. The plane on the right
symbolizes that the solution asymptotes to Euclidean Anti-de
Sitter space.}} \label{fig:non wormhole}
\end{figure}

\subsection{The Sub-Extremal Instanton: $q^2<0$}
Now we deal with solutions which have negative $q^2$. These
solutions share the same symmetry properties as the previous
solutions where $q^2$ was positive, namely they break the
$SO(1,5)$ symmetry of pure EAdS$_5$ down to a $SO(5)$ symmetry.

Defining $\tilde{q}^2=-q^2>0$, the sub-extremal instanton solution
reads:
\begin{equation}\label{sub-extremal solution}
\boxed{\begin{aligned}
ds^2&=\frac{dr^2}{1+\frac{r^2}{l^2}-\frac{\tilde{q}^2}{r^6}}+
r^2d\Omega_{4}^2, \\
e^{\phi(r)}&=|\,\frac{q_-}{\tilde{q}}\sin[\tilde{q}H(r)]\,| , \\
\chi(r)&= \frac{1}{q_- }\Bigl(\tilde{q}
\cot[\tilde{q}H(r)]-q_3\Bigr)\, ,
\end{aligned}}
\end{equation}
where $H(r)$ is an harmonic satisfying equation (\ref{vgl voor
harmonische}) and like in the previous case cannot be obtained
explicitly in D=5, whereas in D=3 explicit results are easy to
obtain (appendix B). Despite this, it is not hard to check (e.g.
numerically) that contrary to the super-extremal instanton, the
harmonic is regular.

The coordinates run from $r=r_c$ to $r=\infty$, where $r_c$ is the
unique root of \begin{equation} 1 + \frac{r^2}{l^2} -
\frac{\tilde{q}^2}{r^6} = 0\,.
\end{equation}
One can check that $r_c$ corresponds to a coordinate singularity
because the Ricci scalar $\mathcal{R}$ stays finite at $r=r_c>0$
since $\mathcal{R}=-12 \tilde{q}^2/r_c^8 - 20/l^2$ . It is
possible to resolve this singularity by making a coordinate
transformation such that the metric takes the following form:
\begin{figure}
\begin{center}
\includegraphics[scale=0.3]{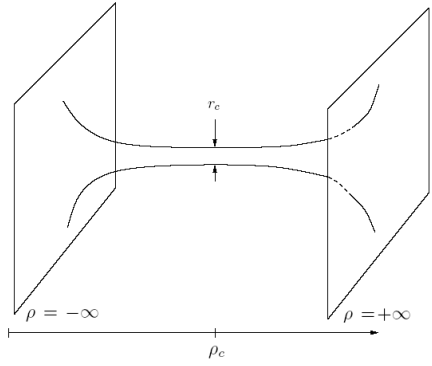}
\end{center}
\caption{\small{For the class $q^2<0$ the space is a wormhole with
the neck of the wormhole at $r=r_c$ and again the planes represent
the fact the geometry asymptotes to pure Euclidean Anti-de Sitter.
}} \label{fig:wormhole}
\end{figure}
\begin{equation} \label{extensie}
ds^2 = d\rho^2 + a(\rho)^2d\Omega_4^2\, .\\
\end{equation}
This implies the relation $a(\rho)=r$. The coordinate $\rho$ goes
from $-\infty$ to $+\infty$, and the function $a(\rho)$ can be
thought of as the scale factor. From the relations between the two
frames we derive an equation for the scale factor \footnote{This
is exactly the same equation found by Gutperle and Sabra in
\cite{Gutperle:2002km}. Their number `$c$' equals
$-24\,\tilde{q}^2$ in our notation.}:
\begin{equation}\label{scalefactor}
(\partial_{\rho} a)^2 = 1 - \frac{\tilde{q}^2}{a^6} +
\frac{a^2}{l^2}.
\end{equation}
The difficulties in $D=5$ encountered in solving this equation
explicitly are the same as solving for the harmonic function for
this space. Although we do not have an explicit solution this is
not a problem since all the necessary information is contained in
this differential equation. The most relevant points that can be
extracted are that this coordinate system `doubles' the original
manifold, and that the scale factor $a(\rho)$ reaches a minimum at
$\rho_c \equiv \rho(r_c)$, and that it is even about this point:
\begin{equation}
a(2\,\rho_c-\rho) = a(\rho)\,.
\end{equation}
Therefore, the sub-extremal D-instanton has the geometry of a
wormhole, with its neck at $\rho_c$ of size $a(\rho_c)$. The
harmonic function $H$ will acquire the following anti-symmetry:
\begin{equation}
H(2\,\rho_c-\rho) = -H(\rho)+2\,H(\rho_c)\,,
\label{Symmetry_harmonic}
\end{equation}
allowing us to easily relate the values of the scalar fields at
both asymptotic regions of the wormhole. We can read of from
(\ref{sub-extremal solution}) that there is a singularity in the
axion and in the derivative of the dilaton whenever
$\tilde{q}H(\rho)$ equals a multiple of $\pi$. If the range
spanned by the image of $\tilde{q}H$ is smaller than $\pi$ we can
always tune $g_s$ such that $\tilde{q}H(\rho)$ never reaches a
multiple value of $\pi$ (i.e the argument of the sine is on the
same branch cut). This range $R$ is given by
$R=|\tilde{q}H(\infty)-\tilde{q}H(-\infty)|$. Using
(\ref{Symmetry_harmonic}) and (\ref{vgl voor harmonische}) we
find:
\begin{equation}\label{range}
R=2\,\sqrt{24}\,\,\tilde{q}\,\int^{+\infty}_{r_c}\frac{1}{r^4\sqrt{1-\frac{\tilde{q}^2}{r^6}+\frac{r^2}{l^2}}}\,dr\,.
\end{equation}
By changing the integration variable to $x=r/l$ one finds that the
integrand and the domain of integration only depend on the combination
$\tilde{q}^2/l^6$ and therefore $R$ only depends on that variable.
Numerical calculations show that there is no value for
$\tilde{q}^2/l^6$ for which $R<\pi$. Hence there is always a
singularity in $\chi$ and in $\phi$, but there is no singularity
in $e^{2\phi}$. It can be shown that, for models with a different
coupling `$b$' of the dilaton to the axion i.e.
\begin{equation}
\mathcal{L_{\text{SCALAR}}}\sim(\partial\phi)^2-e^{b\phi}(\partial\chi)^2\,,
\end{equation}
there exist solutions which are regular in $\chi$ and $\phi$, when
$b<\sqrt{8/3}$. Such values for the dilaton coupling constant could for
instance be obtained from non-spherical compactifications. For the case of
zero cosmological constant, this can be achieved in Calabi-Yau
compactifications of type II strings, as was demonstrated in
\cite{Bergshoeff:2004pg}.

\subsection{Lorentzian Solutions}

Until now we presented three solutions ($q^2>0$, $q^2=0$ and
$q^2<0$) of the Euclidean IIB action involving five-form flux, the
dilaton- and the axion-field. But this subsector of IIB
supergravity also holds interesting solutions of the Lorentzian
theory. One way to try to obtain these solutions is by Wick
rotating the Euclidean solutions and to check whether they are
solutions of the Lorentzian theory. As pointed out in
\cite{Maldacena:2004rf} Euclidean wormholes can Wick rotate to Big
Bang/Big Crunch cosmologies and as we will show this is indeed
what happens in our case. The extremal and super-extremal solution
cannot be Wick rotated to real solutions of Lorentzian IIB
supergravity.

The Wick rotation proceeds by considering the sub-extremal
solution in the $\rho$-coordinates and taking $\rho=i\tau$.  We
can always shift the $\rho$-coordinates such that the neck is
located at the origin. Doing so the Euclidean scale factor
$a_{E}(\rho)$ is an even function and can be Wick rotated to a
real function i.e $\,a_L(\tau)\,:=\,a_E(i\tau)\,\,\in \, \,
\mathbb{R}$. The metric Ansatz becomes the following cosmology:
\begin{equation}
ds^2=-d\tau^2 + a_L(\tau)^2d\Omega_4^2\,.
\end{equation}
The function $a_L(\tau)$ is called the Lorentzian scale factor.
Wick rotating the fields is somewhat more involved. One first has
to notice that (up to an additive constant) $H_E(\rho) \rightarrow
iH_L(\tau)$, where $H_E$ and $H_L$ are the harmonic functions over
the Euclidean and Lorentzian geometry respectively. One has the
freedom to Wick rotate the integration constants as long as one
ends up with a real dilaton field $\phi(\tau)$, and the axion field
$\chi(\tau)$ should get an extra factor of $i$ during the Wick
rotation. These Wick rotation rules are:
\begin{align}\label{Wick rotation rules}
 q_3 & \rightarrow iq_3\, , \\
 H_E & \rightarrow iH_L + \frac{\pi}{2}\, .
\end{align}
The second rule is necessary to keep the dilaton field real and is
to be interpreted as a Wick rotation rule for the integration
constant belonging to the definition of the harmonic function
$H_E$. The solution we end up with reads:
\begin{equation}\label{cosmology}
\boxed{\begin{aligned}
ds^2&=-d\tau^2 + a^2_L(\tau)d\Omega_{4}^2\, , \\
e^{\phi(\tau)}&=\,\frac{|q_-|}{\tilde{q}}\cosh[\tilde{q}H_L(\tau)]\, , \\
\chi(\tau)&= \frac{1}{q_- }\Bigl(\tilde{q}
\tanh[\tilde{q}H_L(\tau)]+q_3\Bigr)\, .
\end{aligned}}
\end{equation}
The $l \rightarrow \infty$ version of this solution was studied in \cite{Lu:2004ms}. To check that this is a solution one has to use the Lorentzian
field equations which give for the scale factor (after combining
the Einstein equation with the scalar field equations) :
\begin{equation} \label{scalefactorII}
(\partial_{\tau} a_L)^2 = -\Bigl[1 - \frac{\tilde{q}^2}{a_L^6} +
\frac{a_L^2}{l^2} \Bigr],
\end{equation}
which differs with an overall minus sign from the equation for the
Euclidean scale factor $a_E(\rho)$ (\ref{scalefactor}). One can
check that the Wick rotated scale factor $a_L(\tau):=a_E(i\tau)$
indeed obeys this equation. The same holds for the scalar fields;
straightforwardly putting them in the Lorentzian field equations
shows that (\ref{cosmology}) indeed is a solution.

This cosmology is a Big Bang/Big Crunch cosmology. This can be
seen either by analyzing the scale factor $a_L(\tau)$ numerically
or by deducing from (\ref{scalefactorII}) that a solution has to
obey
\begin{equation}
  \partial_{\tau}a_L=\left\{ \begin{array}{l}  -\sqrt{-1 + \frac{\tilde{q}^2}{a_L^6}
  -
  \frac{a_L^2}{l^2} }\ \,\,\,\,\quad \text{for}\, \tau > 0,\\
\quad \quad 0 \qquad \qquad \qquad \quad \,\, \text{for} \,  \tau=0,\\
    +\sqrt{-1 + \frac{\tilde{q}^2}{a_L^6} - \frac{a_L^2}{l^2}}\
    \quad \,\,\,\,\, \text{for}\, \tau< 0.
\end{array} \right.
\end{equation}
Between the Big Bang and Big Crunch singularities the scalar
fields are completely regular as one can verify. However, at the
singularities, the harmonic function blows up.
\begin{figure}
\begin{center}
\includegraphics[scale=0.5]{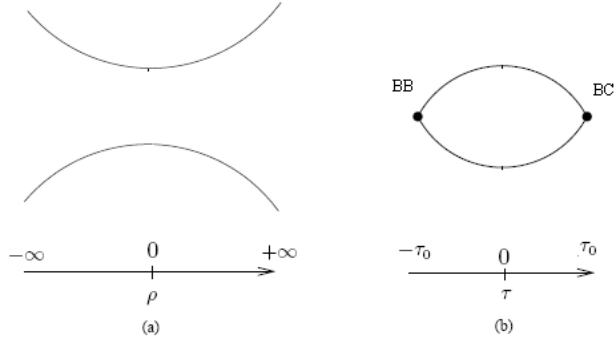}
\end{center}
\caption{\small{Pictorial description of the effect of Wick
rotating a Euclidean wormhole. }} \label{fig:wickwormhole}
\end{figure}

As noted at the end of the section on the super-extremal solution
($q^2>0$), one can take a limit in which the axion becomes a
constant such that the geometry is carried only by a dilaton.
Although this is not possible for the Euclidean sub-extremal
solution, it is possible for the Lorentzian sub-extremal solution
where again the dilaton becomes proportional to the harmonic
function. Such a solution has been discussed in
\cite{Bergshoeff:2005bt}.

\section{Instanton Actions}\label{actions}

As explained in appendix A, in order to calculate the on-shell
instanton action it is convenient to use the Hodge dual formalism
of the axion-dilaton sector of $D=10$ Euclidean IIB supergravity.
This dual theory contains a 9-form $F_9$ field strength instead of
a 1-form field strength $d\chi$. The D-instanton is magnetically
charged with respect to this 9-form. We use the following
Euclidean action:
\begin{equation}
S=-\frac{1}{2\kappa^2_{10}}\int d^{10}x\sqrt{g}\Bigl[
\mathcal{R}-\tfrac{1}{2}(\partial\phi)^2-\tfrac{1}{2}e^{-2\phi}
F_9^2\Bigr]\, .
\end{equation}
Reducing the theory over S$^5$ and using the field equations one
finds:
\begin{equation}\label{vorm_actie}
S = \underbrace{- \frac{1}{2\kappa^2_5}\Bigl(\int_{M}d^5x\,
\sqrt{g}\left(\frac{-8}{l^2}\right)+2\int_{\partial M}d^4x
\sqrt{h}K\Bigr)}_{S_{\text{GRAV}}}
-\underbrace{\frac{1}{2\kappa^2_5}\int_{M}d^5x\,
\partial_{\mu}(\sqrt{g}g^{\mu\nu}\partial_{\nu}
\phi)}_{S_{\text{SCALAR}}}
\end{equation}
We split the action in a part $S_{\text{GRAV}}$ which only
contains the metric and a part only involving the dilaton
$S_{\text{SCALAR}}$. In the gravitational part $S_{\text{GRAV}}$
we have included the usual Gibbons-Hawking term, where
$h_{\mu\nu}$ is the corresponding induced boundary metric
$h_{\mu\nu}=g_{\mu\nu}-n_{\mu}n_{\nu}$ and $K$ is the extrinsic
curvature w.r.t. $h_{\mu\nu}$.  The gravitational part gives an
infinite answer and therefore needs to be regularized. For clarity
we discuss $S_{\text{GRAV}}$ and $S_{\text{SCALAR}}$ separately.

Since $S_{\text{SCALAR}}$ is a total derivative we use Stokes
theorem. After integration over the angles one finds the following
term:
\begin{equation}
S_{\text{SCALAR}}=-\frac{1}{2\kappa_5^2}\text{Vol}(S^4)\eta^r r^4
\partial_{r}\phi \mid_{\partial}\, ,
\end{equation}
where $\mid_{\partial}$ means that the expression is evaluated at
the boundary $\partial$ and $\eta$ is a unit vector perpendicular
to the boundary $\partial$. Using (\ref{vgl voor harmonische})
this can be rewritten as:
\begin{equation}
S_{\text{SCALAR}}=\frac{\sqrt{24}}{2\kappa_5^2}\,\text{Vol}(S^4)\,
e^{-\phi(H)}\partial_{H}e^{\phi(H)}\mid_{\partial}\, .
\end{equation}
This formula is useful because it only requires the knowledge of
the value of $H$ at the boundary and this is completely fixed by
fixing $g_s$ on the boundary. When we have multiple boundaries as
for wormholes we have to subtract the values at different
boundaries \emph{if} the fields are regular everywhere in the
bulk.

If we assume that for the super-extremal instanton, the
singularity gets resolved by string theory effects, then there is
only one boundary at $r=\infty$. This gives:
\begin{equation} \label{action}
S_{\text{SCALAR}}=\frac{\sqrt{6}\,\text{Vol}(S^4)}{\kappa_{5}^2}\,\sqrt{(\frac{q_{-}}{g_s})^2+q^2}\,.
\end{equation}
To obtain the result for the extremal D-instanton we just put
$q^2=0$ in the above expression.

For the sub-extremal solution we find that the singularities in
the fields are not integrable. Hence $S_{\text{SCALAR}}$ formally
diverges. The situation can again be improved by having different
values of the dilaton coupling parameter $b$. As mentioned in the
previous section, this can yield completely regular solutions also
for the scalar fields which will result in a finite-action
instanton. For zero cosmological constant, this was shown in
\cite{Bergshoeff:2004fq}.

Now we focus on $S_{\text{GRAV}}$. The correct way for calculating
the on-shell action for gravitational instantons in an
asymptotically EAdS space requires infrared renormalization
\cite{Emparan:1999pm,Skenderis:1998fq}. The reason is that the
bulk action diverges because of the integration over a non-compact
space. Also the Gibbons-Hawking term diverges since there is a
second order pole in the induced boundary metric $h_{\mu\nu}$.
This means that one has to add counterterms on the regulating
surface $\partial M$ defined by $r=L$. These counterterms cancel
the divergences after taking the regulator $L$ to infinity. The
action with the counterterms is given by:
\begin{equation}\label{full on-shell 5D action}
\begin{split}
S_{\text{GRAV}} =&-\frac{1}{2\kappa_{5}^2}\int_{M}d^5x
\sqrt{g}\Bigl(\frac{-8}{l^2}\Bigr) -\frac{1}{\kappa_{5}^2}
\int_{\partial M}d^4x \sqrt{h}K \\
& +\frac{1}{\kappa_{5}^2}\int_{\partial
M}d^4x\sqrt{h}\Bigl(\frac{3}{l}+\frac{l}{4}\mathcal{R}_h\Bigr)+\frac{1}{\kappa_{5}^2}\int_{\partial
M}d^4x\,\sqrt{h}\, a_4(h) L^4\log{\frac{l}{L}}\, .
\end{split}
\end{equation}
The last counterterm cannot be written in a covariant way and
explicitly contains the regulator. The coefficient $a_4$ is a
covariant function of the induced metric and gives rise to the
conformal anomaly of the dual field theory
\cite{Skenderis:1998fq}. This dual field theory lives on a space
with a metric that is different from the induced metric $h$,
namely the poles have to be dropped such that we get the canonical
S$^4$ metric with radius $l$. If we would have carried out the
same procedure in Poincar\'e coordinates (instead of radial
coordinates) where the regulating surface is defined by $z=1/L$ we
would find that $a_4$ would be zero. The metric on the dual field
theory would be that of $\mathbb{R}^4$ on which $\mathcal{N}=4$
SYM has no conformal anomaly. The meaning of this is that the
regularization procedure picks out a certain induced metric which
will give the metric on which the dual CFT is defined, after the
poles of the induced metric are dropped.

This procedure shows that for the super-extremal solution we get
some extra but finite terms in the action on top of the finite
term generated by pure EAdS$_5$. These extra terms of course
vanish in the limit of $q^2 \rightarrow 0$. The finite term
generated by pure EAdS$_5$ can be put to zero by making an
explicit choice for the scheme dependence of the regularization
procedure. Finally notice that $S_{\text{GRAV}}$ does not depend
on $g_s$, so its contribution does not interfere with the leading
semiclassical (in $g_s$) contribution from $S_{\text{SCALAR}}$.

\section{The ${\rm AdS}_5/{\rm CFT}_4$ Correspondence}\label{curvature}

First we will give some basic facts about the ${\rm AdS}_5/{\rm
CFT}_4$ correspondence which are needed to understand what our
solutions mean in the dual theory. More explanation can be found
in for example \cite{Aharony:1999ti}.

\subsection{Generalities}

On the string theory side we are working in the supergravity
approximation. This means that we assume that $g_sN$ is large,
while $g_s$ stays small, i.e. $1<<g_s N < N $. The number $N$ is
proportional to the 5-form flux and can be seen as the number of
3-branes sourcing this flux. More specific, we have the relations:
\begin{equation}
l^4=Q_3=4\pi g_s N \alpha'^2\, .
\end{equation}
According to the duality $g_{YM}^2 N = 4 \pi N g_s >> 1$ hence we
are dealing with a strongly coupled Yang-Mills theory. The $\rm
AdS/CFT$ duality equates the generating functions of the dual
theories :
\begin{equation}\label{duality}
Z[J=\Phi_{\partial}]=\int\,d[\mathcal{A}]\,\text{exp}\,(-S_{\text{YM}}+J\,\mathcal{O[A]})\,\approx
e^{-S_{\text{SUGRA}}[J]} .
\end{equation}
Here $\mathcal{O[A]}$ represents a gauge invariant SYM operator
which couples to the boundary value of the dual supergravity field
$J=\Phi_{\partial}$. The boundary value $\Phi_{\partial}$ then
clearly acts as a source in the SYM theory. The fact that the
string (supergravity) partition function only depends on
$J=\Phi_{\partial}$ is only correct for bulk configurations which
are sufficiently regular. This duality (\ref{duality}) allows us
to find the 1-point functions of the strongly coupled SYM theory,
since
\begin{equation}\label{duality rule}
-\frac{\delta}{\delta J}\,S_{\text{SUGRA}}[J]= \langle
\,\mathcal{O[A]}\,\rangle_{J}.
\end{equation}
For the dilaton field $\phi$ and the axion field $\chi$ the dual
operators are respectively $\text{Tr}F^2+\ldots$ and
$\text{Tr}(F\tilde{F})$. To be more specific, using the relations
$4\pi g_s=g^2_{YM},\, \chi_{\infty}=\tfrac{\theta_{YM}}{2 \pi}$
and conventions in which the Yang-Mills action looks like
\footnote{These conventions are Tr$(T_aT_b)=-\delta_{ab}$ and
$F_{\mu\nu}^a=\partial_{\mu}A^a_{\nu}-\partial_{\nu}A^a_{\mu}+f_{bc}^a
A^b_{\mu} A^c_{\nu}$.}
\begin{equation}
S_{YM}=-\frac{1}{4{g_{\text{YM}}}^2}\int d^4x
\text{Tr}(F^2)+\ldots
-i\frac{\theta}{32\pi^2}\int d^4x\,\text{Tr}(F\tilde{F})\,
,
\end{equation}
we conclude that:
\begin{align}\label{vev}
&\frac{\delta}{\delta
\phi_{\partial}}S=\frac{1}{4{g_{\text{YM}}}^2}\langle\text{Tr}F^2\rangle +\ldots,  \\
&\frac{\delta}{\delta \chi_{\partial}}S=-\frac{1}{16
\pi}\langle\text{Tr}(F \tilde{F})\rangle,
\end{align}
where the $\ldots$ in the first line can be dropped if no other
fields then the vectors are excited.

In general when studying deformations of AdS$_5 \times$ S$^5$
caused by the backreaction of some matter fields, it affects the
dual description in two possible ways. In one case the deformation
causes the dual field theory to gain certain vacuum expectation
values for Yang-Mills operators which are dual to these matter
fields. This is sometimes referred to as a `vev-deformation'. The
other possibility is an `operator deformation' in which the theory
is changed by adding certain operators to the lagrangian. It can
be argued as follows that the deformations we are considering are
vev-deformations. We mentioned that the boundary value of a bulk
field acts as a source for an operator in the dual field theory,
so generically $\mathcal{N}=4$ SYM  gets deformed by some operator
which couples to that source. However if the operator dual to the
bulk fields is the lagrangian itself the situation is different.
This is of course the case here since the dilaton couples to the
real part of the lagrangian ($\sim
1/g^2_{\text{YM}}$Tr$F^2+\ldots$) and the axion to the imaginary
part ($\sim i\theta_{\text{YM}}$Tr$F\tilde{F}$). The above leads
one to believe that axion-dilaton deformations do not alter the
dual theory but rather pick out a non trivial vacuum (background)
which can spontaneously break (parts of) the conformal symmetry
and supersymmetry. These arguments are not sufficient because they
would imply that all axion-dilaton deformations of (E)AdS (with
undeformed S$^5$) are dual to vev-deformations of $\mathcal{N}=4$
SYM. A well understood example where this is not the case is the
Janus solution \cite{Bak:2003jk}. This is a simple dilatonic
deformation but the dual theory is marginally deformed. There the
reason was that the coupling constant on the boundary makes a
discontinuous step which for instance causes the dual theory to
lose supersymmetry \cite{Clark:2004sb}\footnote{For another
discussion on the holographic dual of Janus, see
\cite{Papadimitriou:2004rz}.}. In our case the dilaton is constant
at the boundary and this phenomenon does not occur, but the
example teaches us that one in general has to be careful.

We shortly give an explanation on how to calculate a variation
with respect to a boundary value in order to calculate the
expectation values. Consider a variation of the action $S=\int
L(\Phi,\partial\Phi)$:
\begin{equation} \delta S= \int [(\frac{\partial L}{\partial \Phi}
-\partial\frac{\partial L}{\partial
\partial\Phi})\delta\Phi+
\partial(\frac{\partial L}{\partial
\partial\Phi}\delta\Phi)]\, .
\end{equation}
The first part is the equation of motion for the field $\Phi$ and
hence disappears on-shell. Using Stokes on the second term we
find:
\begin{equation}\label{variatie} \delta S= \int_{\partial} [
\eta(\frac{\partial L}{\partial
\partial\Phi}\delta\Phi)],
\end{equation}
where $\eta$ is a unit vector perpendicular to the boundary
$\partial$. Expression (\ref{variatie}) is clearly an integral
over the boundary of our space. Hence $\delta \Phi$ in the
integrand can be replaced by its value on the boundary $\delta
\Phi_{\partial}$. From this we read of that,
\begin{equation} \frac{\delta S}{\delta J(\vec{y})}=
\eta_{\mu}\frac{\partial L}{\partial
\partial_{\mu}\Phi}(\vec{y}).
\end{equation}
How to do this properly for $\rm EAdS_5$ is explained for instance
in \cite{Balasubramanian:1998de}. In general one has to use a
regularization technique in order to get finite answers. In
practice, this amounts to making a series expansion of the
supergravity field in the coordinate $z$\cite{Skenderis:2002wp}:
\begin{equation}
\Phi(\vec{y},z)\approx
z^{(4-\Delta)}\Phi(\vec{y})_{\partial}+\ldots +
z^{\Delta}\Phi(\vec{y})_{\,(2\Delta-4)}+\ldots\, ,
\end{equation}
where $\Delta$ is the conformal weight of the dual operator which
is related to the mass of the supergravity field $\Phi$. For
massless fields such as the axion and the dilaton $\Delta=4$.
$\Phi_{\partial}$ represents the source for the dual operator in
the field theory, whereas $\Phi_{\,(2\Delta-4)}$ is related to the
vacuum expectation value of the dual operator.  So the presence of
$\Phi_{\partial}$ in general signals an operator deformation,
unless $\Delta=4$ where the situation is more subtle. In fact, our
deformations are expected to be vev-deformations as explained
above.

\subsection{Calculation of the 1-point Functions} \label{sec:1-pt-fn}

Until now we gave the non-extremal solutions in radial coordinates
but for this section it becomes preferable to present the results
in Poincar\'e coordinates\footnote{The explicit transformation
from radial coordinates to Poincar\'e coordinates is given by
$r(z,\vec{y})=\frac{\sqrt{((l-z)^2+\vec{y}^{\,2})\q((l+z)^2+\vec{y}^{\,2})}}{2z}$.}
since then the coordinates of the dual field theory are the usual
Cartesian coordinates. In the Poincar\'e coordinate system we
approach the boundary by taking small $z$ values and then the
Cartesian $y$-coordinates parameterize the surface approaching the
boundary.

For the $\rm AdS_5/CFT_4$ correspondence one needs to know how the
dilaton and axion behave near the boundary, i.e. near $z=0$. For
the non-extremal solutions there is no explicit expression for the
harmonic but since only the behavior near the boundary is of
importance we can use perturbation theory. For small $z$ we find:
\begin{align}
&\phi(z,\vec{y})\approx\log[{g_s}]+ \frac{8\sqrt{6}l
{\sqrt{q_{-}^2+q^2 {g_s}^{2}}}}{g_s(l^2+\vec{y}^{\,2})^4}\,z^4
+\mathcal{O}(z^6)\ ,\\
&\chi(z,\vec{y})\approx {g_s}^{-1}\sqrt{1+\frac{q^2
{g_s}^2}{{q_{-}^2}}}-\frac{q_3}{ {q_-}} -\frac{8\sqrt{6}l
q_{-}}{(l^2+\vec{y}^{\,2})^4{g_s}^{2}}\, z^4 + \mathcal{O}(z^6)\ .
\end{align}

In order to understand the holographic dual of these instantons we
have to take the quantization of the axion shift symmetry into
account. It will turn out that the dual statement is the fact that
the winding number of the YM instanton is an integer. The Noether
current associated to the shift symmetry of the axion is
\begin{equation}
\mathcal{J_{\mu}}=e^{2\phi}\partial_{\mu}\chi\ .
\end{equation}
Since the exact duality group of type IIB string theory is
expected to be $SL(2, \mathbb{Z})$, the $\mathbb{R}$ shift
symmetry of $\chi$ is broken to a $\mathbb{Z}$ shift symmetry,
i.e. $\chi$ is periodically identified. As we show in appendix
\ref{app_pathint}, this leads to the following quantization of the
axionic charge:
\begin{equation} 2\pi n =
\frac{\pi^3 l^5}{2\kappa_{10}^2}
\int\mathrm{d}\Sigma^{\mu}e^{2\phi}\partial_{\mu}\chi\, ,
\end{equation}
resulting in
\begin{equation}\label{q_- quantization}
q_-=\frac{\sqrt{3/2}\kappa_{10}^2}{4\pi^4 l^5} \, n \, .
\end{equation}
Now we can apply the formula for the 1-point function to the
super-extremal ($q^2>0$) and extremal ($q^2=0$) instanton, to
obtain:
\begin{equation} \label{FF en FF*}
\boxed{\begin{aligned} \langle\text{Tr}F^2
\rangle&=-\frac{192\,l^4}{(l^2+\vec{y}^{\,2})^4}\sqrt{n^2+\frac{32 \pi^8 q^2 g_s^2 l^{10}}{3 \kappa_{10}^4}} ,\\
\langle\text{Tr}F\tilde{F}\rangle&=
-\frac{192\,l^4}{(l^2+\vec{y}^{\,2})^4}\,n\ .
\end{aligned}}
\end{equation}
These expressions show that when $q^2 > 0$ we have a field
strength $F$ which is not (anti-) self-dual.

For the sub-extremal solution we could naively use (\ref{vev}) on
both sides of the wormhole to obtain a result like (\ref{FF en
FF*}) but now for $q^2<0$. That result is simply impossible since
it violates the Cauchy-Schwarz inequality $-\text{Tr}F^2\geq -
\text{Tr}F\tilde{F}$. The reason for this could be twofold; for
multiple boundaries one has to calculate 1-point functions
differently or another reason could be that the singularity in the
axion and dilaton are responsible for this result.

\section{The D-instanton / YM Instanton Correspondence}

In this section we will discuss the correspondence between
D-instantons and Yang-Mills instantons. We explain this
correspondence from the point of view of the expectation values
for the operators Tr$F^2$ and Tr$F\tilde{F}$ as was done for
extremal instantons in \cite{Balasubramanian:1998de}. We first
discuss this for an extremal single-centered D-instanton with
charge $q_{-}$, and then extend this to the non-extremal case
where $q^2>0$. The case of $q^2<0$ is more subtle since the bulk
geometry has two boundaries. Holography with multiple boundaries
is much less well understood; we comment on it at the end of this
section.

\subsection{The extremal case: $q^2=0$ }

{}From (\ref{FF en FF*}) we have for $q^2=0$:
\begin{equation}\label{trF^21}
\langle \text{Tr}F^2
\rangle=-\frac{192\,l^4}{(l^2+\vec{y}^{\,2})^4}\,|n|=\pm \langle
\text{Tr}F\tilde{F} \rangle \, .
\end{equation}
However (\ref{FF en FF*}) can be generalized for extremal
D-instantons since one can place the position at an arbitrary
point of EAdS$_5$ whereas this is not possible for the
non-extremal solutions. If we take the harmonic as in
(\ref{arbitrary-point}) then this results in the more general
expression~\footnote{Strictly speaking, the expressions for
Tr$F^2$ and Tr$F\tilde{F}$ need to be integrated over the collective
coordinates $z_0$ and ${\vec y}_0$ that appear in the instanton
measure in the path integral. Furthermore, the instanton measure also
contains an integration over fermionic collective coordinates. These
need to be saturated by inserting the appropriate number of fermionic
operators. We have suppressed these subtleties here, which also appears on
the supergravity side. For more details on this, see
\cite{Bianchi:1998nk,Dorey:1998xe,Belitsky:2000ws}.}:
\begin{equation}\label{SD-inst}
\langle \text{Tr}F^2
\rangle=-\frac{192\,z_{0}^4}{(z_{0}^2+(\vec{y}-\vec{y_{0}})^2)^4}\,|n|=\pm
\langle \text{Tr}F\tilde{F} \rangle \, .
\end{equation}
For $n=1$ ($n=-1$) this exactly equals the expression for a
classical (anti-)self-dual Yang-Mills instanton with size $z_0$
placed in flat Euclidean space at a position $\vec{y_{0}}$ in
Cartesian coordinates. For $n > 1$, the expression in \eqref{SD-inst} corresponds to a
particular self-dual multi-instanton configuration, which we
specify below. Observe first that this expression is only the
result for the one-point function $\text{Tr}F^2$ in the large $N$
limit of the gauge theory at strong 't Hooft coupling. For large
values of $N$, it was shown in \cite{Dorey:1998qh,Dorey:1999pd}
that to leading order in the saddle-point expansion:
\begin{itemize}
\item The $n$-instanton configuration becomes dominated by $n$
single instantons living in $n$ mutually commuting $SU(2)$
subgroups of $SU(N)$. \item Each of these single instantons are
driven to sit at the same point in moduli space, i.e. their
positions and sizes are equal.
\end{itemize}
The expression in \eqref{SD-inst} is completely consistent with
these results. Based on this, we can schematically write down the
gauge field
\begin{equation} \label{SD-connection}
A_\mu^{SU(N)}(n)= \left( \begin{array}{ccccc}
A_\mu^{SU(2)}(1)  &                     &        & 0 & \\
                   &   A_\mu^{SU(2)}(1)  &        & & \\
                   &                     & \ddots & & \\
               0   &                     &        &   A_\mu^{SU(2)}(1) & \\
                   &                     &        &  &\ddots \\

\end{array} \right) \ ,
\end{equation}
where $A_\mu^{SU(2)}(1)$ stands for a self-dual $SU(2)$ connection
with instanton number one.

Notice that also the values of the action coincide since combining
(\ref{action}) with (\ref{q_- quantization}) we find that (when
$\theta_{\text{YM}}=0$):
\begin{equation}
S_{\text{SUGRA}}=S_{\text{YM}}=\frac{8\pi^2}{g^2_{YM}}\,|n|\ .
\end{equation}
For $\theta\neq0$ the YM action gets a contribution of the form
$i\,\theta\,n$, and in appendix \ref{app_pathint} we will see that
the SUGRA action gets a contribution of the form
$i\,\chi_{\infty}\,q_-$, which can be regarded as the AdS/CFT
partner of the YM topological term, since $\chi_{\infty}$ is
identified with $\theta$ and $q_-$ with $n$. As a final remark we
like to emphasize that the matching of D-instanton configurations
with self-dual Yang-Mills configurations is actually a surprising
result. This is because the D-instanton result yields quantities
in the gauge theory at strong 't Hooft coupling. Instantons in the
gauge theory however, are semiclassical objects that are only
useful in the weakly coupled regime.  The fact that we get results
for the one-point function \eqref{SD-inst} that are easy to
interpret in the weakly coupled gauge theory, hints toward a
non-renormalization theorem that protects the (semi-) classical
value for $\text{Tr}F^2$ from perturbative quantum corrections
\cite{Gopakumar:1999xh}.

\subsection{The non-extremal case: $q^2>0$} \label{interpretation1}

We now consider the case of the super-extremal deformation, with
$q^2>0$. This D-instanton solution is not BPS, and hence we must
be careful interpreting the corresponding operators and
correlation functions on the gauge theory side, since there is no
reason to expect any mechanism that protects quantities from
receiving quantum corrections. The strategy we follow here is to
stay close to the BPS point, and to consider the case where $q^2$
is very small. This is somewhat similar to the description of
near-extremal black holes. In fact, in \cite{Bergshoeff:2004fq} we
showed that when the $q^2>0$ instanton solution in asymptotically
flat space can be uplifted to one higher dimension, it corresponds
to a black hole with charge $Q$ and mass $M$ with $4q^2=M^2-Q^2$.
The near-extremal black hole then indeed corresponds to taking
$q^2<<1$. One might hope that for $q^2$ very small, the result
will not differ too much from the extremal point where the
gauge theory interpretation is well understood in terms of $n$
single Yang-Mills instantons sitting at the same point in moduli
space.

In the presence of a non-vanishing $q^2$, we have seen in
\eqref{FF en FF*} that the result for $\text{Tr}F\tilde{F}$
remains the same whereas $\text{Tr}F^2$ does get deformed.
Consequently, these operators do no longer obey the self-duality
relation as for extremal instantons. Stated differently, the field
strength still has boundary conditions belonging to the same
topological class, labelled by the instanton number $n$, but it
has an anti-self-dual component proportional to $q^2$. Using
\eqref{q_- quantization} as a definition of $n$, the deviation
from self-duality can be written as
\begin{eqnarray} \label{FF min FF*}
\langle\text{Tr}F^2 \rangle -\langle\text{Tr}F\tilde{F}\rangle&=&
-\frac{192 l^4}{(l^2+\vec{y}^{\,2})^4}\,n\,
\Big(\sqrt{1+\frac{q^2g_s^2}{q_-^2}}
-1\Big)\ ,\nonumber\\
&\approx&-\frac{192
l^4}{(l^2+\vec{y}^{\,2})^4}\,n\,q^2\,\Big(\frac{g_s^2}{2q_-^2}
+O(q^2)\Big)\ ,
\end{eqnarray}
where in the second line we have expanded for small $q^2$. This is
the result of the supergravity approximation. We now attempt to
give an interpretation in the gauge theory. As already stressed
before, one must be careful in giving an interpretation in the
weakly coupled gauge theory, since we do not expect that \eqref{FF
min FF*} can be extrapolated from strong to weak 't Hooft
coupling, unless perhaps for very small values of $q^2$. Because
in the non-extremal case there is a (small) anti-self-dual part of
the field strength, it is tempting to associate the deformation
with the presence of anti-instantons. The description of instanton
- anti-instanton configurations in gauge theory is difficult, as
there are no known analytic solutions of the second order
equations of motion that are not self-dual, at least for gauge
groups $SU(2)$ and $SU(3)$. One can work in the dilute gas
approximation where instantons are widely separated, but this
approximation is not valid for the AdS/CFT correspondence, already
in the extremal case. Luckily, for higher rank gauge groups, the
situation simplifies. For large values of $N$ one can still
consider the configuration \eqref{SD-connection}, but we can
deform it by bringing in $SU(2)$ anti-instantons on the diagonal.
Such configurations satisfy the second order equations of motion,
but are not self-dual. The total instanton charge is still given by
$n$, but we distribute it over $k_+$ instantons and $k_-$
anti-instantons with $n=k_+-k_-$ and $k_-$ small:
\begin{equation} \label{inst-connection}
A_\mu^{SU(N)}(n)= \left( \begin{array}{cccc} A_\mu^{SU(2)}(1)  & &
&  0 \cr
                   &  \ddots             &   &  \cr
              &   &   A_\mu^{SU(2)}(1)  &         \cr
      0    &          &       &   A_\mu^{SU(2)}(-k_-)  \cr
\end{array} \right) \ .
\end{equation}
Here $A_\mu^{SU(2)}(-k_-)$ stands for an anti-self-dual $SU(2)$
connection with instanton number $-k_-$. We have taken all the
instantons and anti-instantons at the same point and with the same
size, say at ${\vec y}_0=0$ and $z_0=l$, where the (partial)
annihilation between instantons and the anti-instanton takes
place. One could further distribute the anti-instanton sector into
$k_-$ single charge anti-instantons, but we will not do so in
order to avoid notational complications and because we take $k_-$
small w.r.t. $n=k_+-k_-$ (e.g. $k_-=1$). The presence of an
anti-instanton in a background of instantons of course leads to an
instability. The solution corresponding to (71) is only a saddle
point in the path integral, and the annihilation of anti-instanton
charge will result in perturbative fluctuations around a local
minimum consisting of instanton charge only. We expect a similar
instability on the supergravity side.

With this instanton configuration, we can compute the operators
\begin{eqnarray} \label{YM result}
\text{Tr}F^2 &=& -\frac{192
z_0^4}{(z_0^2+(\vec{y}-\vec{y_0})^2)^4}\, (k_++k_-)
\ ,\nonumber\\
\text{Tr}F\tilde{F}&=& -\frac{192
z_0^4}{(z_0^2+(\vec{y}-\vec{y_0})^2)^4}\, (k_+-k_-)\ .
\end{eqnarray}
Using the fact that $n=k_+-k_-$, for $z_0=l, \vec{y}_0=0$, the operator $\text{Tr}F\tilde{F}$ matches
with the supergravity prediction, while $\text{Tr}F^2$ matches if
we identify
\begin{equation}
\frac{k_-}{n}=\frac{q^2g_s^2}{4q_-^2}\ ,
\end{equation}
to leading order in the small $q^2$ expansion. The left hand side
is a rational number. For this identification to make sense, we
must understand better the quantization condition on $q^2g_s^2$.
This issue can however not be addressed in the supergravity
approximation that we are working in. It would be interesting to
have a better string theory description of the non-extremal
D-instanton where we can address this issue.

\subsection{The non-extremal case: $q^2<0$} \label{interpretation2}

For $q^2=-{\tilde q}^2 < 0$, the bulk space is a (Euclidean) wormhole
with two
disconnected boundaries. Holography and the AdS/CFT correspondence
in the presence of multiple boundaries is not well understood. If
the AdS/CFT correspondence makes sense for such cases, the
conformal field theory lives on the union of the disjoint
boundaries, and so one expects this to be the product of the
theories on the different boundaries. For a recent discussion on
this, see \cite{deBoer:2004yu}. This becomes particularly
problematic for the Euclidean version of the AdS/CFT
correspondence, since then the two boundaries cannot be separated
by horizons that causally separate and disconnect them.

The problem somehow was set aside, after Witten and Yau
\cite{Witten:1999xp} formulated some general criteria under which
(Euclidean) wormholes cannot exist in this context. It applies to
the case of pure gravity when the bulk space is Einstein with negative
cosmological constant and the boundary has non-negative
\footnote{The case of boundaries with negative scalar curvature leads
to instabilities, as was demonstrated in
\cite{Maldacena:2004rf,Buchel:2004rr}. This case seems not to apply in our
situation, although instabilities might still occur. For zero
scalar curvature, the Witten-Yau theorem was shown to hold in
\cite{Cai:2000kd}. Related results are also found or summarized in
\cite{Anderson:2004yi}.} scalar curvature. More recently, however,
Maldacena and Maoz \cite{Maldacena:2004rf} found examples where
wormholes do appear; the Witten-Yau theorem could be avoided by
switching on additional supergravity matter fields, besides only
the metric. For further explanation on how to avoid the Witten-Yau
theorem, see \cite{McInnes:2004ci}, where the case of axionic matter
is discussed. This is precisely the situation we are dealing with.

The wormhole solution we described below suffers from
singularities in the fields. If we for a moment ignore the
difficulty of discussing holography for solutions with singular
fields (by for instance choosing a different value of the dilaton
coupling parameter $b$ ), we can make some remarks. From the
results obtained in section 3, we find that the coupling constant
$g_{YM}$ of the theory would be different on the two boundaries
since:
\begin{equation}
g_s^+=|g_s^-\cos[R]
+\frac{q_{-}}{\tilde{q}}\sin[R]\sqrt{1-\frac{\tilde{q}^2\,g_s^-}{q_{-}^2}}\,\,|
\, ,
\end{equation}
where $g_s^+$ and $g_s^-$ denote the values of the string coupling
on the left $(+)$ boundary and the right $(-)$ boundary
respectively. The parameter $R$ is related to the range of the
harmonic function appearing in the supergravity solution, see
(\ref{range}). This is a bit similar to the dual of the Janus
solution, where also two regions with different coupling constants
appear \cite{Bak:2003jk,Clark:2004sb}. However for the wormhole
the boundaries are disconnected with different couplings on each
side. It seems that, in the approximation we are working, there is
some correlation between the two gauge theories living on the two
boundaries. In a full quantum gravity treatment, where one sums
over all geometries with fixed boundaries, this correlation might
disappear again. For some related discussions, see
\cite{Rey:1998yx,Maldacena:2004rf}.

It seems that the case with $q^2< 0$ has more applications in the
Lorentzian theory, where the solution describes a closed
cosmology. It would be interesting if some version of the AdS/CFT
correspondence can still be applied to this case, in particular to
the physics close to the singularities. We leave this for future
research.

\section{Conclusions}\label{conclusions}

In this paper we have investigated instanton and cosmology
solutions to the (compactified) gravity-axion-dilaton system with
a non-zero cosmological constant. The extremal instanton solutions
have a well-established relation, via the AdS/CFT correspondence,
with the self-dual ${\mathcal N}$=4 Yang-Mills instantons. The
(non-extremal) instanton solutions represent interesting
deformations of the D=5 Euclidean Anti-de Sitter space.

There exist two classes of non-extremal instantons solutions,
depending on the value of the $SL(2,\mathbb{R})$ Noether charge
$q^2$. For the dilaton coupling that we have chosen, only the ones
with $q^2>0$ have a finite action and hence can be considered as
true instantons having a dual description in the ${\mathcal N}=4$
Yang-Mills theory. We have shown that the holographic dual
operators of these non-extremal D-instanton configurations do not
correspond to self-dual Yang-Mills instantons, and we have
computed explicitly the deviation from self-duality in terms of
$q^2$. We have suggested an interpretation of this result as a
specific instanton/anti-instanton configuration where the
different (anti-)instantons take values in mutually commuting
$SU(2)$ subgroups of $SU(N)$.

The other class of solutions, i.e.~the ones with $q^2<0$, only
yield regular and finite action solutions for restricted values of
the dilaton coupling. They describe Euclidean wormhole solutions
which, after an appropriate Wick rotation,  correspond to exact
closed cosmology solutions. We hope that these Euclidean wormhole
solutions allow for a holographic description of cosmological
singularities.

\section*{Acknowledgments}
We would like to thank Ioannis Papadimitriou, Stefano Kovacs,
Kasper Peeters, Marija Zamaklar, Miranda C.N. Cheng and Elisabetta
Pallante for discussions. S.V. is supported by the European
Commission FP6 program MRTN-CT-2004-005104 while E.B., A.C., A.P.
and T.V.R. are supported by the  European Commission FP6 program
MRTN-CT-2004-005104 in which E.B., A.C., A.P. and T.V.R. are
associated to Utrecht University. The work of A.P. and T.V.R. is
part of the research programme of the ``Stichting voor
Fundamenteel Onderzoek der Materie'' (FOM).

\appendix
\section{Path integral formulation of axion-dilaton gravity} \label{app_pathint}
In this appendix, we will establish the path integral formulation
of axion-dilaton gravity that leads to D-instanton solutions in
the semiclassical approximation. Through this formulation, we will
be able to explain the `wrong' sign of the axionic kinetic term in
\eqref{D=5_Bulkactie} and the presence of non-gravitational
boundary terms needed to properly evaluate the actions of our
solutions. As a bonus, we will see how the analog of the
Yang-Mills $\theta$-term arises on the gravity side. This
discussion is based on explanations found in \cite{Brown:1989df,
Burgess:1989da, Coleman:1989zu} and references therein.
\bigskip

In Euclidean field theory, one is usually interested in computing
the partition function:
\begin{equation} Z=\langle\phi_F, \chi_F |
e^{-H\,T} | \phi_I, \chi_I \rangle, \quad {\rm in \ the \ limit}
\quad T \rightarrow \infty\,,
\end{equation}
where $\phi_{I, F}$ and $\chi_{I,F}$ are the dilaton and axion
evaluated at the initial and final spacelike\footnote{Note that in
order to give an instanton interpretation to the solutions in this
paper, one must not choose `$r$' as the Euclidean `time' direction
since RR-charge is conserved with respect to it. A good candidate
is the $y_0$-direction in Poincar\'e coordinates.} surfaces
$\Sigma_{I,F}$. This can be written in path integral language as
follows:
\begin{equation} Z = \int_{\rm b.c.}
d[\phi]\,d[\chi]\,\exp \left[-\tfrac{1}{2} \int_{\mathcal{M}} \,
(d\phi \wedge \ast d\phi + e^{b\,\phi}\, d \chi \wedge \ast d\chi)
\right]\,, \label{pathint1}
\end{equation}
where Dirichlet boundary conditions are imposed on all fields. For
practical purposes, we will omit the gravitational sector in this
discussion, as it is not relevant to this discussion. We will also
temporarily omit the integration over the dilaton and its kinetic
term to keep formulae short and will reinsert everything when we
are finished. Although formula \eqref{pathint1} gives us in
principle all the information we need, it can only be computed in
the semiclassical approximation, where instanton contributions
will be highly suppressed and sub-leading compared to perturbation
theory terms. Therefore, in order to see any instanton effects, it
is useful to reorganize this calculation by inserting complete
sets of momentum eigenstates of the axion \cite{Coleman:1989zu}.
The latter are defined as follows:
\begin{equation}
|\pi\rangle \equiv \int d[\chi]
\exp\left(i\,\int_{\Sigma}\,\pi\,\chi\right)\,|\chi\rangle\,,
\end{equation}
where the integral is a functional integral over $\chi$, and $\pi$
is the `timelike' component of a one-form (i.e. transverse to
$\Sigma$). This is completely analogous to the relation between
momentum and position eigenstates in quantum mechanics. Inserting
two complete sets of momentum states at $t=t_I$ and $t=t_F$, we
rewrite our path integral as follows:
\begin{align}
Z &= \int d[\pi_I]\,d[\pi_F] \langle\chi_F | \pi_F\rangle\ \langle
\pi_F |\, e^{-H\,T} |\, \pi_I\rangle \langle \pi_I | \chi_I
\rangle \nonumber \\
&= \int d[\pi_I]\,d[\pi_F] \exp\left(i\,\int_{\Sigma_F}
\pi_F\,\chi_F-i\,\int_{\Sigma_I}
\pi_I\,\chi_I\right)\,\langle\pi_F |\, e^{-H\,T} |\,
\pi_I\rangle\,. \label{pathint2}
\end{align}
We will discuss the interpretation of the surface term we just
generated later on. In \eqref{pathint2}, we have Fourier
transformed the boundary data of our path integral with respect to
the axion field. Instead of computing an amplitude between two
axion field states $\chi_I$ and $\chi_F$, we now have to compute
an amplitude between two momentum states $\pi_I$ and $\pi_F$. To
do so, we need to Fourier transform the initial and final states
back to the original field variables.
\begin{equation} K_E (\pi_F,
\pi_I, T) \equiv  \langle\pi_F |\, e^{-H\,T} |\, \pi_I\rangle =
\int d[\bar{\chi}_I]\,d[\bar{\chi}_F] \langle\pi_F |
\bar{\chi}_F\rangle \langle\bar{\chi}_F |\, e^{-H\,T} |\,
\bar{\chi}_I \rangle \langle\bar{\chi}_I | \pi_I\rangle\,,
\end{equation}
where $\bar{\chi}_I$ and $\bar{\chi}_F$ should be thought of as
`dummy' variables that have nothing to do with the boundary
conditions $\chi_I$ and $\chi_F$ in our original path integral
\eqref{pathint1}. If we now write $\langle\bar{\chi}_F |\,
e^{-H\,T} |\, \bar{\chi}_I \rangle $ as a path integral, with
$\bar{\chi}_{I,F}$ as boundary conditions, and combine the
integration over the bulk field $\chi$ with the integrations over
$\bar{\chi}_{I, F}$, we are left with the following path integral,
which has no boundary conditions:
\begin{equation} K_E = \int_{\rm no \ b.c} d[\chi]\,\exp
\left(-\tfrac{1}{2} \int_{\mathcal{M}} \, e^{b\,\phi}\, d \chi
\wedge \ast d\chi-i\,\int_{\Sigma_F}
\pi_F\,\chi+i\,\int_{\Sigma_I} \pi_I\,\chi \right)\,.
\label{pathint3}
\end{equation}
This is the path integral we would like to approximate. Note that
the boundary term here has nothing to do with the boundary term in
\eqref{pathint2}, because the boundary values of $\chi$ here are
`dummy' variables that we are integrating over, whereas in
\eqref{pathint2}, they are the true boundary conditions of the
original problem. Instanton effects contributing to $K_E$ will
then be interpreted as tunnelling processes that cause axionic
charge conservation violation. However, if we try to compute $K_E$
via the standard saddle point approximation, we immediately run
into a contradiction. Varying w.r.t. $\chi$, we find the following
two equations:
\begin{equation} d \left( e^{b\,\phi}\,\ast d\chi
\right)=0\, \qquad {\rm and} \quad \left.e^{b\,\phi}\,\ast
d\chi-i\,\pi \right|_{\Sigma_{I, F}}=0\,.
\end{equation}
The first looks like a normal bulk equation of motion. The second
equation, however, is a boundary term, which we cannot throw away
because no boundary conditions have been imposed in our path
integral. This equation contradicts the assumption that $\chi$ and
$\pi$ are real. Hence, we conclude that there are no non-trivial
real saddle points for \eqref{pathint3}. Therefore, we need to
find a different method to approximate $K_E$. We will now show
that we can rewrite the path integral for $K_E$ in terms of a
\emph{dual} variable to $\chi$, namely a $(D-1)$-form. This dual
formalism will allow us to perform a saddle point approximation
for \eqref{pathint3}. Let us define the dual path integral as
follows:
\begin{equation} \int d[F]\,d[\chi]\,\exp
\left(\int_{\mathcal{M}} -\tfrac{1}{2}\, e^{-b\,\phi}\,F \wedge
\ast F + i\,\chi\,dF\right)\,, \label{pathint4}
\end{equation}
where $F$ is a $(D-1)$-form. We impose the following Dirichlet
boundary conditions on the `magnetic' part of $F$, (i.e.~on the
`spacelike' components, which are the components along
$\Sigma_{I,F}$):
\begin{equation} F_{||,I,
F} = \ast \pi_{I, F}\,,
\end{equation}
and we impose no boundary conditions on $\chi$. Through a simple
integration by parts, a shifting of variables and a Gaussian
integration, we can easily eliminate $F$ from \eqref{pathint4},
and the result will be \eqref{pathint3}. This is analogous to the
path integral in quantum mechanics. Usually, one derives from
first principles a path integral where both $x$ and $p$ are
variables of integration, and then one eliminates $p$ in favor of
$x$. In this case, however, we want to compute an amplitude
between momentum eigenstates, therefore, it is more natural to
eliminate the `position' variable in favor of the momentum,
i.e.~eliminate $\chi$ in favor of $F$. The integral over $\chi$ in
\eqref{pathint4} yields a $\delta$-functional: $\delta[dF]$. This
imposes the constraint that $F$ be closed. Neglecting global
issues for simplicity, this means that we can write $F$ as a
field-strength $F=dC_{D-2}$. Then it is easy to perform the
semiclassical approximation for this system. We can derive the
following equation of motion:
\begin{equation} d(e^{-b\,\phi}\,
\ast F)=0\,,
\end{equation}
which means that, locally, one can rewrite the field-strength as
follows:
\begin{equation} F = e^{b\,\phi} \ast d\lambda\,,
\label{onshellduality}
\end{equation}
where $\lambda$ is a scalar. The equation of motion of the dilaton
is the following:
\begin{equation} d\ast
d\phi+\frac{b}{2}\,e^{-b\,\phi}\,F\wedge \ast F = 0\,.
\end{equation}
Substituting the definition of $\lambda$ into this yields the
following:
\begin{equation} d\ast d\phi+\frac{b}{2}\,e^{b\phi}\,d\lambda
\wedge \ast d\lambda=0\,.
\end{equation}
This equation of motion has the wrong sign in front of the
$\lambda$ term. One can similarly show that the Einstein equation
also `sees' an axion with the wrong sign. Hence, the remaining
equations of motion of the resulting system are the ones we have
been solving in this chapter; i.e.~those of a system with a wrong
sign kinetic term for the axion. At the end of the day, the result
of solving the $F$ equations and substituting the solution into
\eqref{pathint4} is effectively the same as performing a saddle
point approximation of a `would-be' imaginary scalar field
$\lambda$ with the following action:
\begin{equation}
S=\int_{\mathcal{M}} \tfrac{1}{2}\, \left[ d\phi \wedge \ast d\phi
- e^{b\,\phi}\, d \lambda \wedge \ast d\lambda +2\,d
\left(\lambda\,e^{b\,\phi}\, \ast d\lambda \right)\right]\,,
\label{wouldbe}
\end{equation}
and with the following Neumann boundary conditions for the axion
current:
\begin{equation} \left.e^{b\,\phi}\,d\lambda
\right|_{\Sigma_{I, F}} = \pi_{I, F}\,.
\end{equation}
To summarize, we start by setting up a path integral for a
transition between eigenstates of the conjugate momentum to the
axion field $\chi$, and we generate a boundary term that implies
that no real saddle points can be found. By studying the system in
its dual formulation, we are able to perform the semiclassical
approximation and find that the result can be effectively obtained
by defining an imaginary axion field with Neumann boundary
conditions. Let us now reinsert $K_E$ into our original path
integral \eqref{pathint2}, and elucidate the nature of the
remaining boundary term:
\begin{equation} Z = \langle\chi_I |
e^{-H\,T} | \chi_F\rangle = \int d[\pi_I]\,d[\pi_F]
\exp\left(i\,\int_{\Sigma_F} \chi_F\, \pi_F-i\,\int_{\Sigma_I}
\chi_I \,\pi_I\right)\,K_E(\pi_I, \pi_F, T)\,.
\end{equation}
If we restrict the path integral to axion fields with equal
constant parts at $\Sigma_{I}$ and $\Sigma_{F}$, and we call that
part $\chi_{\infty}$, then the surface term has the following
contribution:
\begin{equation} i\,\chi_{\infty}\,\Delta Q\,, \qquad
{\rm where} \quad \Delta Q \equiv \int_{\Sigma_F}
\pi_F-\int_{\Sigma_I} \pi_I \label{theta}\, .
\end{equation}
In terms of the parameters in the solutions we discuss in this
paper, $\Delta Q$ corresponds to $q_-$. In the AdS/CFT dictionary,
$\chi_{\infty}$ translates to the vacuum angle $\theta$ in
super-Yang-Mills, and, as we saw in section \ref{sec:1-pt-fn},
$q_-$ translates to $k$, the instanton number. Hence, this term in
the SUGRA action seems to be the AdS/CFT partner of the
topological term $i\,\theta\,k$ in super-Yang-Mills. Since the
invariance of the action under constant $\mathbb{R}$-shifts of the
axion is expected to be broken by string theory effects to
$\mathbb{Z}$-shift invariance, the appearance of this contribution
\eqref{theta} in our SUGRA action implies that $q_-$, or $\Delta
Q$ is quantized. The argument is as follows: $\chi$ is
periodically identified. Hence, in order to have a single valued
path integral, $\Delta Q$ must be quantized, which is consistent
with the AdS/CFT relation between $k$ and $\Delta Q$.

\section{Explicit Solutions in $D=3$}
In $D=3$ the harmonic function is a compact expression and hence
things can be shown more explicitly compared to $D=5$, it can also
be useful for ${\rm AdS_3/CFT_2}$. This system can be seen as a
compactification of IIB supergravity on a 7-dimensional space of
the form ${\rm AdS_3 \times S^3} \times M_4$, where $M_4$ is a
compact hyperKh\"aler manifold ($T^4$ or $K^3$). The dilaton
coupling $b$ is left as a free parameter, since its value may
differ from two via the above mentioned compactification. The
\emph{super}-extremal solution reads:
\begin{equation}\label{solutionsblobalads3}
\begin{split} &
ds^2=\frac{dr^2}{1+\frac{r^2}{l^2}+\frac{q^2}{r^{2}}}+
r^2d\Omega_{2}^2\, ,\\ & H(r)=\frac{ \text{arcsinh}[\frac{ q
g_s^{b/2}}{q_-}]}{q}-\frac{b \log[2+\frac{l}{q}]}{2q} + \frac{b
\log[\frac{l}{q}+\frac{2ql}{r^2}+\frac{2\sqrt{r^4+(q^2+r^2)l^2}}{r^2}]}{2q}\, ,\\
& e^{b\phi}=\Bigl( \frac{q_{-}}{q}\sinh[qH]\Bigr)^2\, ,\\
& \chi=\frac{2}{bq_{-}}\Bigl( q\coth[qH]-q_{3} \Bigr).
\end{split}
\end{equation}
The \emph{sub}-extremal solution ($-q^2=\tilde{q}^2>0$) reads:
\begin{equation}
\begin{split} &a(\rho)^2=\frac{1}{2}(-l^2+\sqrt{4l^2\tilde{q}^2+l^4}\cosh[\frac{2\rho}{l}]), \\
& H(\rho)=\frac{\arcsin[\frac{\tilde{q}
g_s^{b/2}}{q_-}]}{\tilde{q}}+
\frac{b}{\tilde{q}}\text{arctan}[\frac{l^2+\sqrt{4 \tilde{q}^2l^2
+l^4}}{2\tilde{q}l}]-
\frac{b}{\tilde{q}}\text{arctan}[\frac{(l^2+\sqrt{4 \tilde{q}^2l^2 +l^4})\tanh[\frac{\rho}{l}]}{2\tilde{q}l}]\, . \\
\end{split}
\end{equation}
We skipped the expressions for the dilaton and axion since they
differ similarly from the super-extremal solution as in $D=5$. The
neck of the wormhole is at $\rho=0$. This solution can be Wick
rotated, resulting in:
\begin{equation}
\begin{split} a(\tau)^2&=\frac{1}{2}(-l^2+\sqrt{4 l^2 \tilde{q}^2+l^4}\cos[\frac{2 \tau}{l}]) , \\
H(\tau)&=\frac{\arcsin[\frac{\tilde{q}
g_s^{b/2}}{q_-}]}{\tilde{q}}+
\frac{b}{\tilde{q}}\text{arctan}[\frac{l^2+\sqrt{4 \tilde{q}^2l^2
+l^4}}{2\tilde{q}l}]\\& \quad  -
\frac{b}{\tilde{q}}\text{arctanh}[\frac{(l^2+\sqrt{4
\tilde{q}^2l^2 +l^4} )\tan[\frac{\tau}{l}]}{2\tilde{q}l}].
\end{split}
\end{equation}
The axion can be truncated as follows:
\begin{equation}
H(\tau)\rightarrow H(\tau) -\frac{\arcsin[\frac{q
g_s^{b/2}}{q_-}]}{q}+ \frac{\text{arcsinh}[\frac{q
g_s^{b/2}}{q_-}]}{q} , \quad\quad
q_3\rightarrow -q + \frac{b q_+ q_-}{2q}\, ,
\end{equation}
after which one takes the limit $q_-$ to zero to obtain a
solution.
\bibliography{ADSinstantons}
\bibliographystyle{utphysmodb}
\end{document}